\newcommand{\PRE}[1]{}       % Use if journal style
\documentclass[aps,prl,twocolumn,preprintnumbers,superscriptaddress,
              showpacs,nofootinbib]{revtex4-1}
\usepackage{bm}
\usepackage{graphicx}

\newcommand{\gev}{\text{GeV}}

\newcommand{\pb}{\text{pb}}

\newcommand{\cm}{\text{cm}}
\newcommand{\m}{\text{m}}
\newcommand{\km}{\text{km}}

\newcommand{\s}{\text{s}}

\newcommand{\eqref}[1]{Eq.~(\ref{#1})}

\newcommand{\sigmaSI}{\sigma_{\rm SI}}
\newcommand{\sigmaSD}{\sigma_{\rm SD}}

\newcommand{\be}{\begin{equation}}
\newcommand{\ee}{\end{equation}}
\newcommand{\ssection}[1]{{\em #1.}}

\newcommand{\bea}{\begin{eqnarray}}
\newcommand{\eea}{\end{eqnarray}}

\newcommand{\lsim}{\lower.7ex\hbox{$\;\stackrel{\textstyle<}{\sim}\;$}}
\newcommand{\gsim}{\lower.7ex\hbox{$\;\stackrel{\textstyle>}{\sim}\;$}}

\newcommand{\nbox}{{\,\lower0.9pt\vbox{\hrule \hbox{\vrule height 0.2 cm
\hskip 0.2 cm \vrule height 0.2 cm}\hrule}\,}}

\begin{document}

\preprint{UH-511-1165-2011}

\title{ \PRE{\vspace*{1.5in}}
Dark Matter Detection With Electron Neutrinos in Liquid
Scintillation Detectors
\PRE{\vspace*{0.3in}} }

\author{Jason Kumar, John G.~Learned, Michinari Sakai and Stefanie Smith \\%
%\PRE{\vspace*{.2in}}
Department of Physics and Astronomy, University of
Hawai'i, Honolulu, HI 96822, USA
%\PRE{\vspace*{.2in}}
}

%\date{August 2008}

\begin{abstract}

\PRE{\vspace*{.3in}} We consider the prospects for liquid scintillation
experiments (with a focus on KamLAND) to detect
the flux of electron neutrinos arising from dark matter annihilation
in the core of the sun.  We show that, with data already taken, KamLAND
can provide the greatest sensitivity to the dark matter-proton spin-dependent
scattering cross-section for dark matter lighter than 20 GeV.  It is also
possible to probe the dark matter-nucleon spin-independent scattering
cross-section for isospin-violating dark matter lighter than 10 GeV.  KamLAND can
thus potentially confirm the dark matter interpretation of the
DAMA and CoGeNT signals, utilizing data already taken.

\end{abstract}

%, 04.65.+e, 12.60.Jv}
%95.35.+d Dark matter
%04.65.+e Supergravity
%12.60.Jv Supersymmetric models
%95.55.Vj Neutrino, muon, pion and other elementary particle detectors

\maketitle

\ssection{Introduction}
Neutrino detectors search for the neutrino flux arising from dark matter annihilating in the sun's
core.
It has been argued recently that liquid scintillator (LS) neutrino detectors can
be used for dark matter searches~\cite{Learned:2009rv,Kumar:2009ws}.
The key is the ability of LS detectors to reconstruct a charged
lepton track from the timing of the first scintillation photons to reach the PMTs.  If a
charged lepton is produced from a neutrino from the sun through a charged-current interaction,
then a measurement of the direction and energy of the fully-contained charged lepton track is
sufficient to reconstruct the neutrino energy.  The measured energy spectrum can
then be compared to the expected atmospheric neutrino background.

This analysis is typically performed utilizing muon tracks, where the track direction
is determined from the Cerenkov cone.  The difficulty with this method is
that, unless the muons are of relatively low energy or the detector is extremely large, the
muon track will not be contained within the detector.  This makes it impractical to measure the
energy of the muon, and thus impossible to determine the energy of the original $\nu_\mu$.
Instead, one must compare the event rate of muons which pass entirely through the
detector (``throughgoing muons") to the event rate expected from atmospheric neutrinos.  Since
the atmospheric neutrino background falls sharply with energy, the throughgoing muon background
is dominated by low-energy neutrinos which produce muons just energetic enough to pass through
the fiducial volume.

There are two significant advantages to dark matter searches for $\nu_e$, $\bar \nu_e$, producing
$e^-$ or $e^+$ via charged-current interactions.  First, the atmospheric neutrino flux for
electron neutrinos is smaller than that of $\nu_\mu$, $\bar \nu_\mu$ by a factor which varies
from $\sim 2$ to $\sim 10$ in the energy range of interest.  More importantly,
unlike muons, electrons and positrons produce showers which attenuate very quickly.  Even a
very energetic $\nu_e$ will produce a shower which can be fully-contained within a reasonably
sized LS detector.  For such a shower, the timing of the first detected photons can be
used to reconstruct the direction of the produced electron/positron with $\lesssim 1^\circ$
uncertainty~\cite{Learned:2009rv}.  With the direction and energy of the electron/positron,
as well as total calorimetry, one can reconstruct the energy of the neutrino to within
$\lesssim 1\%$~\cite{Learned:2009rv}.
On the other hand, water Cerenkov (WC) detectors such as Super-Kamiokande have greater difficulty in accurately
measuring the direction or energy of electron showers, making it difficult for them to base
dark matter searches on electron neutrinos.

This suggests that electron neutrinos are an ideal channel for neutrino-based dark matter searches,
and are the channel for which LS detectors are uniquely well-suited.  We will demonstrate that KamLAND,
using data already collected, can place bounds on the spin-dependent dark matter-proton scattering
cross-section ($\sigmaSD^p$) which are competitive with current bounds.  We will also show that
KamLAND can probe the dark matter-nucleon spin-independent scattering cross-section ($\sigmaSI$)
at a level competitive with other experiments for $m_X \lsim 10~\gev$.

\ssection{Dark Matter Detection with $\nu_e ,\bar \nu_e$}
Dark matter is gravitationally captured by the sun through elastic scattering from solar nuclei: when dark
matter loses enough energy through nuclear recoil, it falls below the sun's escape velocity
and is captured, eventually settling in the core.
The capture rate depends on the dark-matter-nucleon scattering cross-section ($\sigma_{XN}$),
the dark matter mass ($m_X$), the local dark matter density and velocity distribution,
and on the composition of the sun~\cite{Gould:1987ir,Jungman:1995df}.

For dark matter in the range of masses
considered here, the sun would be in equilibrium~\cite{equilib}, with the capture rate $\Gamma_C$ related
to the annihilation rate $\Gamma_A$ by $\Gamma_C = 2\Gamma_A$.  Given any choice of the
dark matter annihilation channel, $\Gamma_A$ determines the magnitude of the neutrino flux at earth, while
$m_X$ determines the neutrino energy spectrum.  These together determine
the lepton interaction rate at any neutrino detector.  Since $\Gamma_C$ is determined
by $m_X$ and $\sigma_{XN}$, a measured event
rate at a neutrino detector constrains the $(m_X, \sigma_{XN})$ parameter-space.

\ssection{KamLAND}
KamLAND is an LS detector with an approximately spherical inner detector
($V \sim 1000~\m^3$).  The KamLAND scintillator density is
$\sim 80\%$ that of water.  We consider a ``fully-contained" electron
event to be an $e^+$ or $e^-$ shower which starts within the detector and travels for
at least 4.3 m before leaving the inner detector.  This corresponds to $\sim 10$ radiation
lengths, ensuring a light yield sufficient to accurately determine the energy of the
$e^+$, $e^-$ initiating the shower.  The volume for this analysis is
the portion of the inner detector in which a track pointing from the sun could originate and
travel at least 4.3 m without leaving the inner detector.  Our analysis volume is $\sim {1\over 2}$
the volume of the inner detector.

\ssection{Analysis}
The fully-contained charged lepton rate at a neutrino detector can be written as
\bea
R_{l(\bar l)} &=& \Gamma_A \times {\sigma_{\nu (\bar \nu) N} (m_X) \times N_A \over 4\pi R^2}
\times \langle Nz \rangle_{\nu (\bar \nu)}
\eea
where $z= E_{\nu} / m_X$, $N_A$ is the number of target nucleons within the analysis volume,
$R=1.5 \times 10^{11}~\m$ is the earth-sun
distance, and $\sigma_{\nu (\bar \nu) N} $ are the (anti-)neutrino-nucleon
scattering cross-sections.  In the range $E_{\nu} \sim 2-1000~\gev$, these can be approximated
as~\cite{Edsjo:1997hp,Gandhi:1995tf}
\bea
\sigma_{\nu N} (E_\nu) &=& 6.66 \times 10^{-3}~\pb \left( E_\nu / \gev \right)
\nonumber\\
\sigma_{\bar \nu N} (E_{\bar \nu}) &=& 3.25 \times 10^{-3}~\pb \left( E_{\bar \nu} / \gev \right)
\eea
This cross-section is thus proportional to $\langle Nz \rangle_{\nu (\bar \nu)}$, where
\bea
\langle Nz \rangle = {1\over m_X} \int_0^{m_X} dE \, \left[ {dN \over dE} E \right]
\eea
is the first moment
of the neutrino spectrum.

$E_\nu$ is determined by the electron energy ($E_e$) and the angle $\theta$ between
the electron shower and the neutrino:
\bea
E_\nu &=& m_N E_e [ m_N -E_e (1-\cos \theta)]^{-1},
\eea
where $m_N$ is the nucleon mass.
Since KamLAND can measure the energy and direction of fully-contained leptons precisely, it can determine
$E_\nu$ event by event for a neutrino assumed to be arriving from the direction of the sun.
Our analysis counts only events where the electron shower is
within an analysis cone of half-angle
\bea
\label{coneEQ}
\theta_{cone} = 20^\circ \sqrt{10~\gev / E_\nu}
\eea
from the direction from the sun; $\sim 2/3$ of electrons arising from the
charged-current interaction of an
electron-neutrino originating in the sun will lie within this cone.

Note that this analysis can be refined significantly if the direction and energy
of the recoiling nucleon can also be measured from scintillation light, as has been
argued in~\cite{Learned:2009rv}.  This measurement would permit the neutrino energy and
direction to be reconstructed independently, greatly reducing the background from
atmospheric neutrinos.  However, for an analysis at KamLAND, the number of background
events is so small that this further step is not required.

The atmospheric electron neutrino flux can be determined from Honda {\it et al.}~\cite{Honda:2011nf}.  For a search for
dark matter with mass $m_X$, we count events with reconstructed $E_\nu$ between
$E_{thr}=1.5~\gev$ and $m_X$; for a 2135 live-day search for $m_X = 5-1000~\gev$, there will be
fewer than 5 $e^\pm$ events arising from atmospheric neutrinos within the analysis cone given in
eq.~\ref{coneEQ} (averaging over zenith angle, azimuthal
angle and solar cycle).
KamLAND can be considered to be sensitive to models which would produce 10 signal events
in 2135 live-days.

$\Gamma_C$ and $\langle Nz \rangle$ are determined as functions of
$m_X$ and $\sigma_{XN}$ by DarkSUSY~\cite{Gondolo:2004sc}, including the effects of neutrino
oscillations in vacuum and matter~\cite{MSW,Pakvasa:2004hu} on the neutrino spectra.
Dark matter local density is taken to be $0.3~\gev /\cm^3$ with a Maxwell-Boltzmann
velocity distribution with dispersion $\bar v = 270~\km / \s$.
The neutrino spectra have been computed for $b$, $\tau$, $W$ and $\nu_{e, \mu ,\tau}$ annihilation channels.
It is assumed that the $W^\pm$ polarization is isotropic.  If the dark
matter is a Majorana fermion, then $W$'s will be transversely polarized~\cite{Drees:1992am}.  However,
the assumption of an isotropic polarization will have only a negligible effect on $\langle Nz \rangle$,
assuming $E_{thr} \ll m_X$~\cite{Jungman:1994jr}.

\begin{figure}[tb]
\includegraphics[width=0.95\columnwidth]{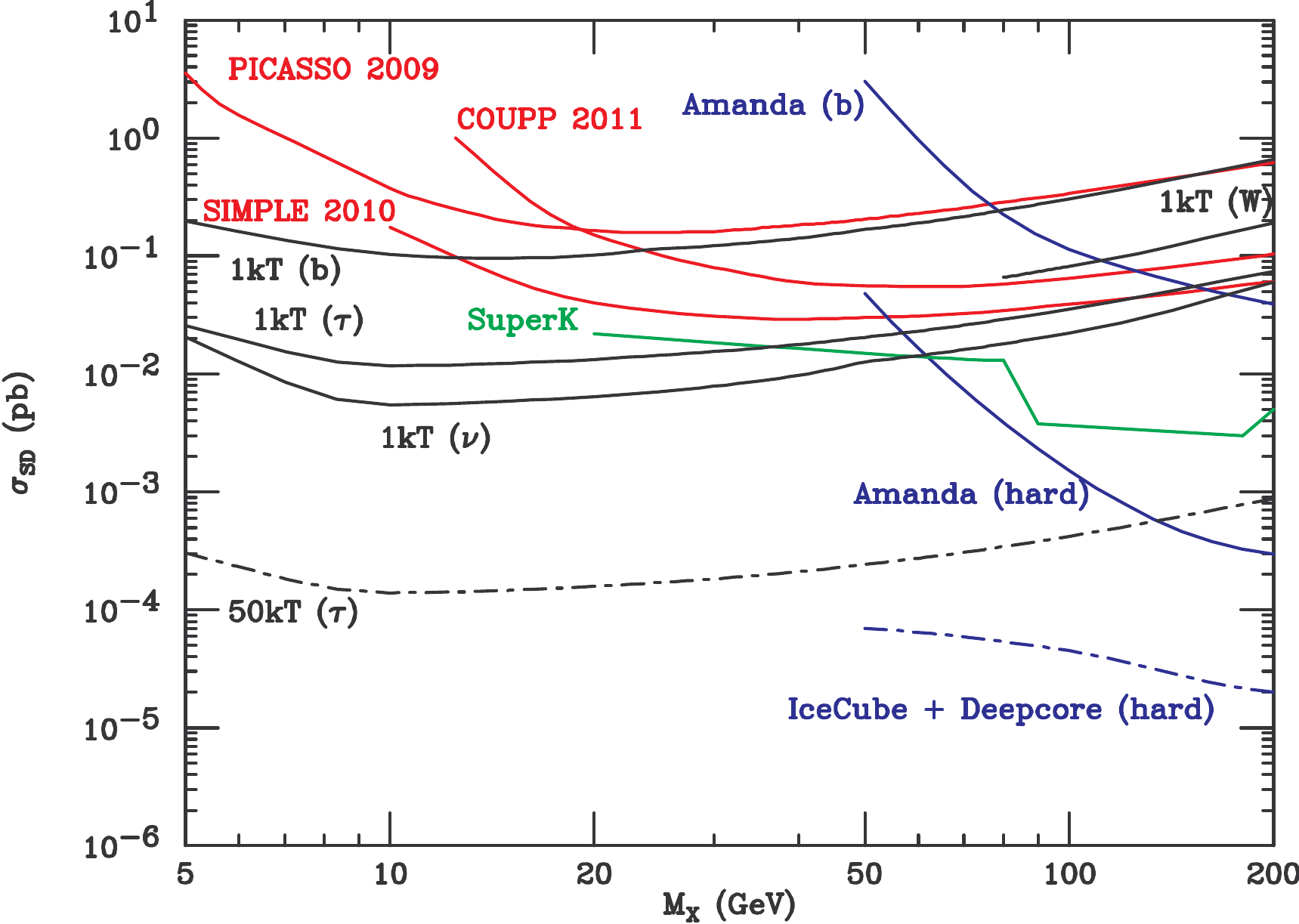}
\includegraphics[width=0.95\columnwidth]{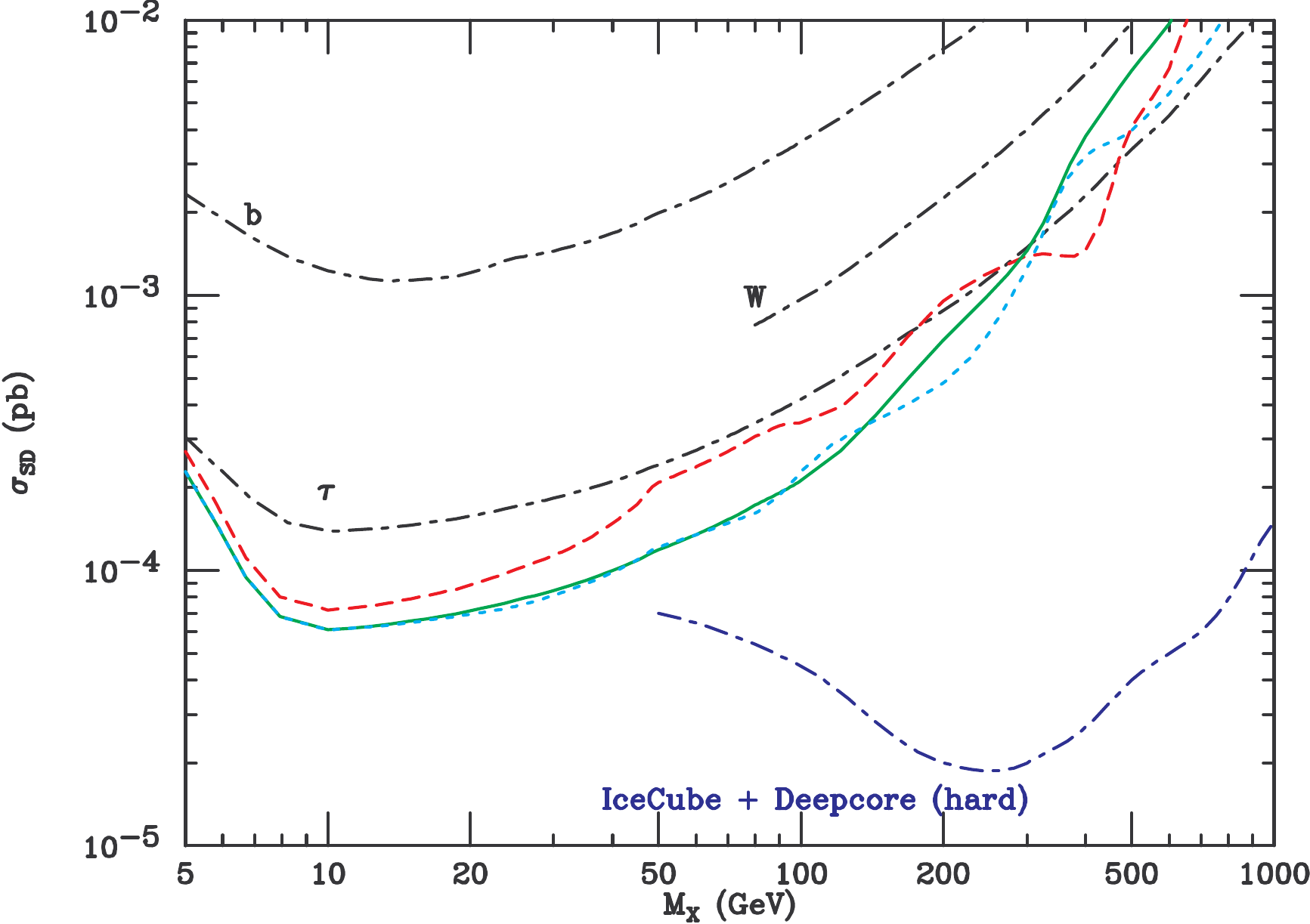}
\vspace*{-.1in}
\caption{\label{fig:SDbounds} (Top panel) Sensitivity of a 1 kT LS detector (such as
KamLAND) to $\sigmaSD^p$, using
2135 live-days of data, assuming annihilation to the $b$, $\tau$, $W$ or $\nu$ (flavor-blind) as
labeled.  Also plotted are current bounds from Super-Kamiokande~\cite{Desai:2004pq},
PICASSO~\cite{Archambault:2009sm}, COUPP~\cite{Behnke:2010xt}, SIMPLE~\cite{SIMPLE},
Amanda and IceCube, as well as prospective bounds from IceCube/DeepCore with
1800 live-days of data~\cite{Braun:2009fr} and prospective bounds for a 50 kT LS detector with
1800 live-days of data.
The hard channel for Amanda and IceCube is the $\tau$ channel for $m_X < 80~\gev$ and the
$W$ channel for $m_X > 80~\gev$.
(Bottom panel) Sensitivity of a 50 kT LS detector with 1800 live-days of data, assuming annihilation
to the $b$, $\tau$, $W$ and $\nu_e$ (dashed), $\nu_\mu$ (solid) and $\nu_\tau$ (dotted) channels as labeled.}
\end{figure}

\ssection{Bounds from KamLAND}
Fig.~\ref{fig:SDbounds} shows the sensitivity to $\sigmaSD^p$ which KamLAND can achieve assuming
2135 live-days of data and dark matter annihilation entirely to the $b$, $\tau$, $W$ or
$\nu$ (flavor-blind) channels.  This bound arises
from dark matter captured by the sun through spin-dependent scattering off hydrogen.  Also reported
in Fig.~\ref{fig:SDbounds} are bounds on $\sigmaSD^p$ from PICASSO, COUPP, SIMPLE,
Super-Kamiokande (Super-K), Amanda,
and IceCube/DeepCore and a projection for a 50 kT future LS detector
(e.~g.~ LENA~\cite{Marrodan Undagoitia:2006re} or Hanohano~\cite{Learned:2007zz}).
The 50 kT LS detector projection assumes 1800 live-days of data.

Even for $m_X > 80~\gev$, detection prospects for annihilation
through the $\tau$ channel are better than for the $W$ channel.
This is not the case for Amanda or IceCube, due to several effects.  The hardest $\nu$
spectrum arises from transversely-polarized $W$ bosons, which
are heavily peaked at large and small values of $z$.  For detectors searching for
fully-contained muons, the event rate is proportional to $\langle Nz^2 \rangle$, thus
weighting transversely polarized $W$s more heavily.  This is especially true for detectors
whose energy threshold is comparable to the dark matter mass; since they are only sensitive
to $\nu$s with $z> E_{thr}/m_X$, the best detection prospects arise from spectra peaked at
large $z$.  For KamLAND, the analysis threshold
is always much lower than $m_X$, and the event rate for fully-contained
events is proportional to $\langle Nz \rangle$.

Note that, if dark matter couples to quarks through heavy mediators, then an effective
operator analysis can permit the Tevatron to place current exclusion bounds in the
$\sigmaSD^p \sim 10^{-3}-10^{-4}~\pb$ range for $m_X \lsim 10~\gev$~\cite{Goodman:2010yf}.

\begin{figure}[tb]
\includegraphics[width=0.95\columnwidth]{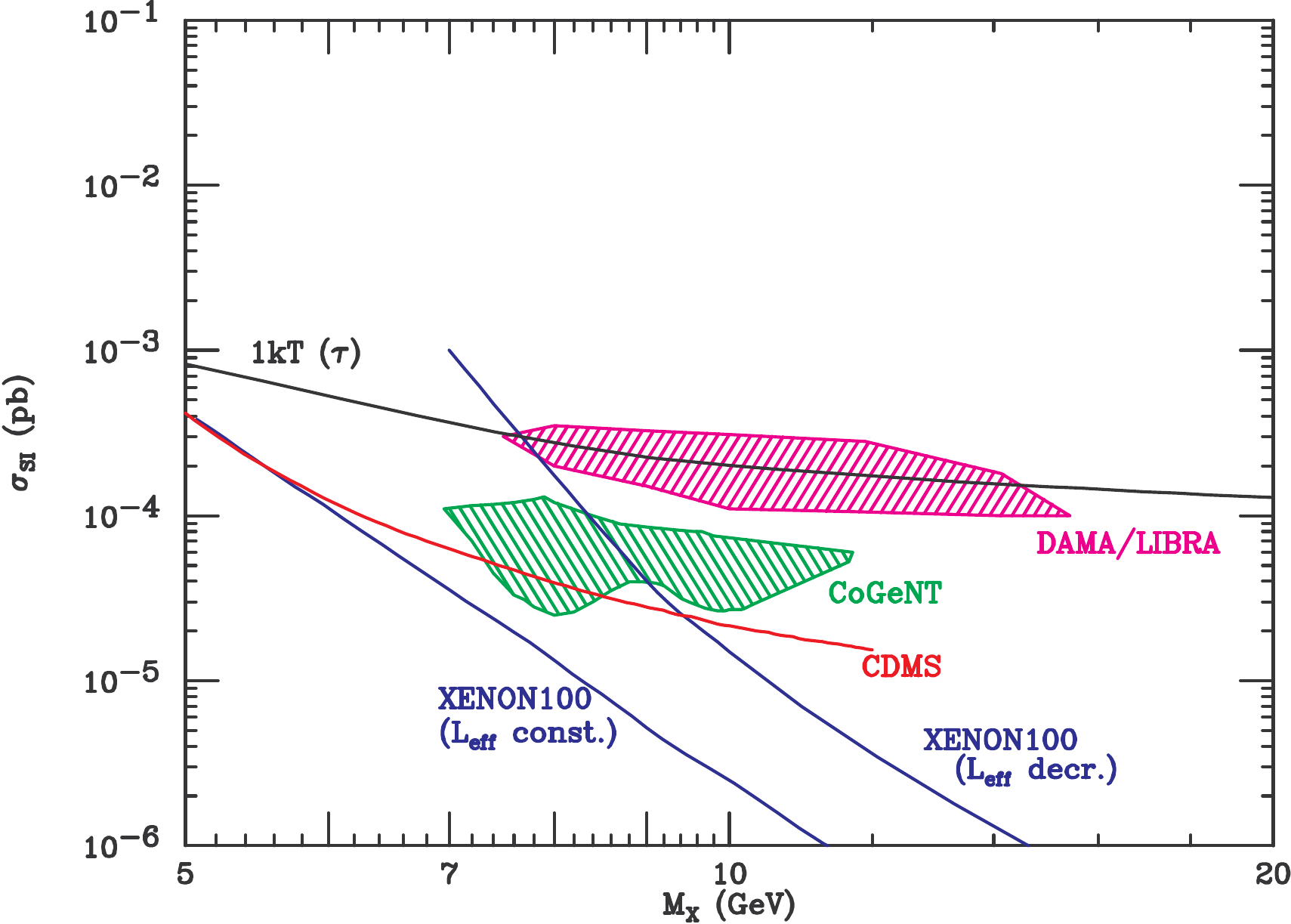}
\includegraphics[width=0.95\columnwidth]{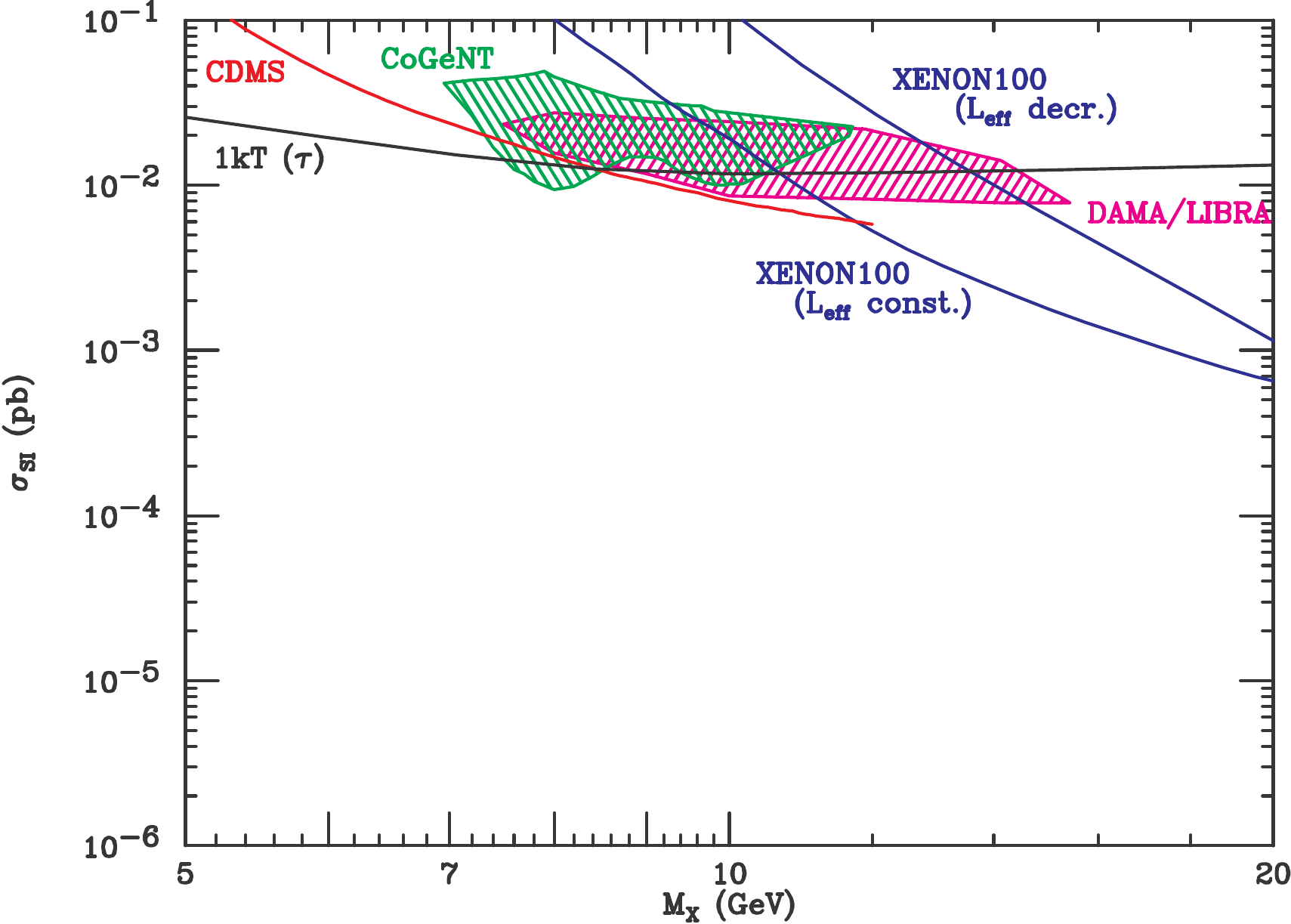}
\vspace*{-.1in}
\caption{\label{fig:SIbounds}  Sensitivity of a 1 kT LS detector (such as KamLAND)
to $\sigmaSI^N$ (black), using
2135 live-days of data and assuming annihilation in the $\tau$ channel.
Also plotted are the preferred region for
the CoGeNT signal (green) at 90\% CL~\cite{Aalseth:2010vx}, the preferred
DAMA region (magenta) at
$3\sigma$ CL (no channeling)~\cite{Savage:2008er,Savage:2010tg}, and exclusion bounds from
CDMS Soudan (red)~\cite{Akerib:2010pv} and Xenon100 (blue)~\cite{Aprile:2010um}, with either
${\cal L}_{eff.}$ either constant or decreasing with energy. The top panel assumes
no isospin violation ($f_n = f_p$).  The bottom panel is for an IVDM model
with $f_n / f_p = -0.7$.}
\end{figure}

Fig.~\ref{fig:SIbounds} shows the sensitivity to the dark matter-nucleon spin-independent scattering
cross-section ($\sigmaSI^N$) which KamLAND can achieve with
2135 live-days of data and dark matter annihilation
to $\tau$s.  The spin-independent (SI) capture rate is
dominated by scattering off heavier nuclei; though heavy nuclei are rare in the
sun, dark matter-nucleus scattering receives an $A^2$ coherent scattering enhancement.
Bounds on $\sigmaSI$ are thus tighter than those on $\sigmaSD^p$.  But
since direct detection experiments are so much more sensitive to $\sigmaSI$, the bounds from
KamLAND are only relevant for $m_X \lsim 10~\gev$, when direct detection experiments begin to
lose sensitivity.

This region of parameter-space is especially interesting, since the DAMA and CoGeNT experiments
have reported signals which are potentially consistent with a dark matter candidate with $m_X \lsim 10~\gev$
and $\sigmaSI \sim 10^{-3-5}~\pb$~\cite{Bernabei:2010mq,Aalseth:2010vx}.  CRESST has also reported preliminary data~\cite{CRESST}
which is consistent with the DAMA and CoGeNT signals.  However, exclusion bounds from the XENON100~\cite{Aprile:2010um}
and CDMS~\cite{Akerib:2010pv} collaborations are in tension with a dark matter interpretation of DAMA, CoGeNT and
CRESST.  Reanalyses of Xenon10 data are also in tension with these signals~\cite{Sorensen:2010hv,Savage:2010tg}.
There is much controversy regarding both the reported signals and the exclusion bounds,
in particular regarding the sensitivity of these direct detection experiments at low mass~\cite{lowmassquestions}.
There is thus great interest in testing these results with a different experimental method.
Super-K can potentially probe
this region of parameter-space with data already taken~\cite{Feng:2008qn,Kumar:2009ws} utilizing muon tracks,
though this analysis of the data has not yet been performed.  Assuming equal dark matter
couplings to protons and neutrons, KamLAND can probe part of the DAMA-preferred region, but not CoGeNT
(top panel, Fig.~\ref{fig:SIbounds}).

But it has recently been noticed that the data from DAMA and CoGeNT and the bounds from CDMS and
Xenon10/100 can be brought into better agreement if one considers isospin-violating dark
matter (IVDM)~\cite{Chang:2010yk,Feng:2011vu}.  IVDM couples
differently to protons and neutrons; if we parameterize these couplings by $f_{p,n}$, the data seem to be
brought into closest agreement for $f_n /f_p \sim -0.7$.  Since dark matter coupling to protons and neutrons
interfere destructively, direct detection experiments which rely on coherent scattering suffer a great loss of
sensitivity.  But for $m_X \sim 10~\gev$, $\sim 3\%$ of dark matter capture is due to
scattering from hydrogen~\cite{Zentner:2009is}, where there is no destructive interference.  Thus, KamLAND may be more
sensitive to IVDM models which can explain DAMA and CoGeNT than are other direct detection experiments.

A conservative estimate of KamLAND's sensitivity
is to assume that IVDM only scatters against the hydrogen.
The sensitivity of KamLAND to $\sigmaSI^p$ in
this limit is the same as to $\sigmaSD^p$.  Taking $f_n / f_p \sim -0.7$,
DAMA and CoGeNT could be consistent with an IVDM particle with a SI cross-section for scattering
off a proton given by $\sigmaSI^p \sim 2-3 \times 10^{-2}~\pb$ for $m_X \sim 10~\gev$~\cite{Feng:2011vu}.
If the IVDM candidate has a significant annihilation branching fraction to $\tau$'s, it can be probed by
data already taken at KamLAND (see bottom panel of Fig.~\ref{fig:SIbounds}).

\ssection{Comparison to Water Cerenkov Detectors}
Water Cerenkov neutrino detectors can also search for electron neutrinos
produced by dark matter annihilating in the sun.  However, for large exposures,
LS detectors are expected to have much smaller backgrounds.

In addition to precise measurement of the charged lepton energy and direction,
LS detectors can independently measure the neutrino energy to
within $\sim 1\%$ accuracy~\cite{Learned:2009rv}
using the total light yield (including scintillation from the recoiling nucleon).
The direction of the electron neutrino can thus be reconstructed to within
$1^\circ$ accuracy~\cite{Learned:2009rv}.
This analysis should be able to reject most atmospheric background within the analysis cone of
eq.~\ref{coneEQ}, leaving only background events from atmospheric neutrinos arriving within $\sim 1^\circ$
of the sun.  This method of background rejection cannot
be used by WC detectors, which cannot independently measure the neutrino energy and can
only reconstruct it under the assumption the neutrino came from the sun.

As a specific example, we may compare the sensitivity of a 50 kT LS detector to a
putative 200 kT WC detector (the estimated size of LBNE~\cite{LBNE}), assuming a run-time of
1800 live-days. Given the larger
exposure, one would expect $\sim 1144$ electron events due to atmospheric neutrinos
within the analysis cone (eq.~\ref{coneEQ}) at the WC detector.
The WC detector can be considered sensitive ($2\sigma$) to dark matter models producing $\sim 68$
signal events within the cone over this run-time.  For the 50 kT
LS detector, assuming $1^\circ$ uncertainty in neutrino direction resolution, one would expect
less than 2 background events; this detector would be sensitive to models which would
produce $\sim 10$ signal events over the run-time.  Accounting for the larger volume of
the WC detector, one would still expect the LS detector's sensitivity to be greater than the WC detector's
sensitivity by a factor $\sim 1.7$.  It should be emphasized, however, that the ability to
independently reconstruct the neutrino direction depends crucially on the ability to measure the
scintillation light of the recoiling nucleon; the efficiencies and uncertainties in the measurement
must be understood for a designed detector before firm conclusions about sensitivity can be made.

\ssection{Conclusion}
We have studied the dark matter detection prospects for KamLAND, using the 2135 live days
of running which are already available.
KamLAND can provide the world's best sensitivity to
the $\sigmaSD^p$ for $m_X \sim 4-20~\gev$.
Moreover, KamLAND's sensitivity to dark matter is not as heavily suppressed by isospin-violating
destructive interference as that of other direct detection experiments for $m_X \sim 10~\gev$.
If the $\tau$ annihilation channel dominates, KamLAND's sensitivity to IVDM is
competitive with other direct
detection experiments, and can potentially test recent hints of low-mass dark matter from
DAMA, CoGeNT and CRESST.  Though KamLAND is a smaller detector than Super-K, this
disadvantage is compensated by the ability to search for dark matter in the
$\nu_e$, $\bar \nu_e$ channel.

Recently, it was argued that a light dark matter candidate which could potentially explain
the DAMA and CoGeNT signals could also be responsible for a possible photon excess
from the Galactic Center, provided the candidate annihilates primarily to $\tau$'s~\cite{Hooper:2010mq}.
LS detectors can thus provide an interesting way of testing this model.

Future large LS detectors such as Hanohano and LENA can improve sensitivity by perhaps $\times 100$.
But a complete analysis must include a simulation of acceptances and efficiencies
of a particular detector, including energy and angular resolution, $e/\mu$ discrimination, nucleon
recoil measurement,
and cosmic ray $\mu$ rejection.  These issues are beyond the scope of this work.

\ssection{Acknowledgments}
We are grateful to M.~Drees, S.~Dye, M.~Felizardo, P.~Gondolo and D.~Marfatia for useful discussions.
This work is supported in part by Department of Energy grant
DE-FG02-04ER41291.


\begin{thebibliography}{99}

%\cite{Learned:2009rv}
\bibitem{Learned:2009rv}
  J.~G.~Learned,
  %``High Energy Neutrino Physics with Liquid Scintillation Detectors,''
  arXiv:0902.4009 [hep-ex];
  %%CITATION = ARXIV:0902.4009;%%
  %\cite{Peltoniemi:2009xx}
%\bibitem{Peltoniemi:2009xx}
  J.~Peltoniemi,
  %``Liquid scintillator as tracking detector for high-energy events,''
  [arXiv:0909.4974 [physics.ins-det]].
  %\cite{Peltoniemi:2009hv}


%\cite{Kumar:2009ws}
\bibitem{Kumar:2009ws}
  J.~Kumar, J.~G.~Learned, S.~Smith,
  %``Light Dark Matter Detection Prospects at Neutrino Experiments,''
  Phys.\ Rev.\  {\bf D80}, 113002 (2009).
  [arXiv:0908.1768 [hep-ph]].

%\cite{Gould:1987ir}
\bibitem{Gould:1987ir}
  A.~Gould,
  %``Resonant Enhancements in WIMP Capture by the Earth,''
  Astrophys.\ J.\  {\bf 321}, 571 (1987).

\bibitem{Jungman:1995df}
  G.~Jungman, M.~Kamionkowski and K.~Griest,
  %``Supersymmetric dark matter,''
  Phys.\ Rept.\  {\bf 267}, 195 (1996)
  [arXiv:hep-ph/9506380].
  %%CITATION = PRPLC,267,195;%%


\bibitem{equilib}
%\bibitem{Gould:1987ju}
  A.~Gould,
  %``WIMP DISTRIBUTION IN AND EVAPORATION FROM THE SUN,''
  Astrophys.\ J.\  {\bf 321}, 560 (1987);
  %%CITATION = ASJOA,321,560;%%
%\bibitem{Griest:1986yu}
  K.~Griest and D.~Seckel,
  %``Cosmic Asymmetry, Neutrinos and the Sun,''
  Nucl.\ Phys.\  B {\bf 283}, 681 (1987)
  [Erratum-ibid.\  B {\bf 296}, 1034 (1988)];
  %%CITATION = NUPHA,B283,681;%%
%\bibitem{Kamionkowski:1994dp}
  M.~Kamionkowski {\it et al.}
  %, K.~Griest, G.~Jungman and B.~Sadoulet,
  %``Model independent comparison of direct versus indirect detection of
  %supersymmetric dark matter,''
  Phys.\ Rev.\ Lett.\  {\bf 74}, 5174 (1995)
  [arXiv:hep-ph/9412213].
  %%CITATION = PRLTA,74,5174;%%
 %\cite{Gould:1987ju}

  %\cite{Gandhi:1995tf}
\bibitem{Gandhi:1995tf}
  R.~Gandhi {\it et al.}
  %, C.~Quigg, M.~H.~Reno and I.~Sarcevic,
  %``Ultrahigh-energy neutrino interactions,''
  Astropart.\ Phys.\  {\bf 5}, 81 (1996)
  [arXiv:hep-ph/9512364];
  %%CITATION = APHYE,5,81;%%
%\cite{Hisano:2009fb}
%\bibitem{Hisano:2009fb}
  J.~Hisano, K.~Nakayama and M.~J.~S.~Yang,
  %``Upward muon signals at neutrino detectors as a probe of dark matter
  %properties,''
  Phys.\ Lett.\  B {\bf 678}, 101 (2009)
  [arXiv:0905.2075 [hep-ph]];
  %%CITATION = PHLTA,B678,101;%%
  %\cite{Erkoca:2009by}
%\bibitem{Erkoca:2009by}
  A.~E.~Erkoca, M.~H.~Reno and I.~Sarcevic,
  %``Muon Fluxes From Dark Matter Annihilation,''
  arXiv:0906.4364 [hep-ph].
  %%CITATION = ARXIV:0906.4364;%%

%\cite{Edsjo:1997hp}
\bibitem{Edsjo:1997hp}
  J.~Edsjo,
  %``Aspects of neutrino detection of neutralino dark matter,''
  arXiv:hep-ph/9704384.
  %%CITATION = HEP-PH/9704384;%%

%\cite{Honda:2011nf}
\bibitem{Honda:2011nf}
  M.~Honda {\it et al.}
  %, T.~Kajita, K.~Kasahara and S.~Midorikawa,
  %``Improvement of low energy atmospheric neutrino flux calculation using the
  %JAM nuclear interaction model,''
  arXiv:1102.2688 [astro-ph.HE].
  %%CITATION = ARXIV:1102.2688;%%

%\cite{Gondolo:2004sc}
\bibitem{Gondolo:2004sc}
  P.~Gondolo {\it et al.}
  %, J.~Edsjo, P.~Ullio, L.~Bergstrom, M.~Schelke, E.~A.~Baltz,
  %``DarkSUSY: Computing supersymmetric dark matter properties numerically,''
  JCAP {\bf 0407}, 008 (2004)
  [astro-ph/0406204].

\bibitem{MSW}
%\cite{Wolfenstein:1977ue}
%\bibitem{Wolfenstein:1977ue}
  L.~Wolfenstein,
  %``Neutrino oscillations in matter,''
  Phys.\ Rev.\  D {\bf 17}, 2369 (1978);
  %%CITATION = PHRVA,D17,2369;%%
%\cite{Mikheyev:1989dy}
%\bibitem{Mikheyev:1989dy}
  S.~P.~Mikheyev, A.~Y.~Smirnov,
  %``Resonant neutrino oscillations in matter,''
  Prog.\ Part.\ Nucl.\ Phys.\  {\bf 23}, 41 (1989);
  %%CITATION = PPNPD,23,41;%%
N.~Itoh, {\it et al.} 1996, ApJS, 102:411-424.


%\cite{Pakvasa:2004hu}
\bibitem{Pakvasa:2004hu}
  S.~Pakvasa,
  %``Neutrino properties from high energy astrophysical neutrinos,''
  Mod.\ Phys.\ Lett.\  A {\bf 19}, 1163 (2004)
  [Yad.\ Fiz.\  {\bf 67}, 1179 (2004)]
  [arXiv:hep-ph/0405179].
  %%CITATION = YAFIA,67,1179;%%


%\cite{Drees:1992am}
\bibitem{Drees:1992am}
  M.~Drees, M.~M.~Nojiri,
  %``The Neutralino relic density in minimal N=1 supergravity,''
  Phys.\ Rev.\  {\bf D47}, 376-408 (1993).
  [hep-ph/9207234].

%\cite{Jungman:1994jr}
\bibitem{Jungman:1994jr}
  G.~Jungman, M.~Kamionkowski,
  %``Neutrinos from particle decay in the sun and earth,''
  Phys.\ Rev.\  {\bf D51}, 328-340 (1995).
  [hep-ph/9407351].

%\cite{Marrodan Undagoitia:2006re}
\bibitem{Marrodan Undagoitia:2006re}
  T.~Marrodan Undagoitia {\it et al.}
  %, F.~von Feilitzsch, M.~Goder-Neff, K.~A.~Hochmuth, L.~Oberauer, W.~Potzel and M.~Wurm,
  %``Low energy neutrino astronomy with the large liquid scintillation detector
  %LENA,''
  Prog.\ Part.\ Nucl.\ Phys.\  {\bf 57}, 283 (2006)
  [J.\ Phys.\ Conf.\ Ser.\  {\bf 39}, 278 (2006)]
  [arXiv:hep-ph/0605229].
  %%CITATION = 00462,39,278;%%

%\cite{Learned:2007zz}
\bibitem{Learned:2007zz}
  J.~G.~Learned, S.~T.~Dye and S.~Pakvasa,
  %``Hanohano: A Deep ocean anti-neutrino detector for unique neutrino physics
  %and geophysics studies,''
  arXiv:0810.4975 [hep-ex].
  %%CITATION = ARXIV:0810.4975;%%



\bibitem{Desai:2004pq}
  S.~Desai {\it et al.}  [Super-Kamiokande Collaboration],
  %``Search for dark matter WIMPs using upward through-going muons in
  %Super-Kamiokande,''
  Phys.\ Rev.\  D {\bf 70}, 083523 (2004)
  [Erratum-ibid.\  D {\bf 70}, 109901 (2004)]
  [arXiv:hep-ex/0404025].
  %%CITATION = PHRVA,D70,083523;%%

  %\cite{Archambault:2009sm}
\bibitem{Archambault:2009sm}
  S.~Archambault {\it et al.}
  %, F.~Aubin, M.~Auger, E.~Behnke, B.~Beltran, K.~Clark, X.~Dai, A.~Davour {\it et al.},
  %``Dark Matter Spin-Dependent Limits for WIMP Interactions on F-19 by PICASSO,''
  Phys.\ Lett.\  {\bf B682}, 185-192 (2009).
  [arXiv:0907.0307 [hep-ex]].


%\cite{Behnke:2010xt}
\bibitem{Behnke:2010xt}
  E.~Behnke {\it et al.}
  %, J.~Behnke, S.~J.~Brice, D.~Broemmelsiek, J.~I.~Collar, P.~S.~Cooper, M.~Crisler, C.~E.~Dahl {\it et al.},
  %``Improved Limits on Spin-Dependent WIMP-Proton Interactions from a Two Liter CF$_3$I Bubble Chamber,''
  Phys.\ Rev.\ Lett.\  {\bf 106}, 021303 (2011).
  [arXiv:1008.3518 [astro-ph.CO]].

\bibitem{SIMPLE}
%\cite{Felizardo:2010mi}
%\bibitem{Felizardo:2010mi}
  M.~Felizardo,
  %T.~Morlat, A.~C.~Fernandes, T.~A.~Girard, J.~G.~Marques, A.~R.~Ramos, M.~Auguste, D.~Boyer
  {\it et al.},
  %``First Results of the Phase II SIMPLE Dark Matter Search,''
  Phys.\ Rev.\ Lett.\  {\bf 105}, 211301 (2010).
  [arXiv:1003.2987 [astro-ph.CO]];
%\cite{Girard:2011xc}
%\bibitem{Girard:2011xc}
  T.~Girard {\it et al.} [ for the SIMPLE Collaboration ],
  %``New limits on WIMP interactions from the SIMPLE dark matter search,''
  [arXiv:1101.1885 [astro-ph.CO]].





%\cite{Braun:2009fr}
\bibitem{Braun:2009fr}
  J.~Braun and D.~Hubert for the IceCube Collaboration,
  %``Searches for WIMP Dark Matter from the Sun with AMANDA,''
  [arXiv:0906.1615 [astro-ph.HE]].

%\cite{Goodman:2010yf}
\bibitem{Goodman:2010yf}
  J.~Goodman, {\it et al.},
  %M.~Ibe, A.~Rajaraman, W.~Shepherd, T.~M.~P.~Tait, H.~-B.~Yu,
  %``Constraints on Light Majorana dark Matter from Colliders,''
  Phys.\ Lett.\  {\bf B695}, 185-188 (2011).
  [arXiv:1005.1286 [hep-ph]];
  %\cite{Goodman:2010ku}
%\bibitem{Goodman:2010ku}
  J.~Goodman, {\it et al.},
  %M.~Ibe, A.~Rajaraman, W.~Shepherd, T.~M.~P.~Tait, H.~-B.~Yu,
  %``Constraints on Dark Matter from Colliders,''
  Phys.\ Rev.\  {\bf D82}, 116010 (2010).
  [arXiv:1008.1783 [hep-ph]].




\bibitem{Bernabei:2010mq}
  R.~Bernabei {\it et al.},
  %``New results from DAMA/LIBRA,''
  arXiv:1002.1028 [astro-ph.GA].
  %%CITATION = ARXIV:1002.1028;%%

%\cite{Aalseth:2010vx}
\bibitem{Aalseth:2010vx}
  C.~E.~Aalseth {\it et al.}  [CoGeNT collaboration],
  %``Results from a Search for Light-Mass Dark Matter with a P-type Point
  %Contact Germanium Detector,''
  arXiv:1002.4703 [astro-ph.CO];
  %%CITATION = ARXIV:1002.4703;%%
  %\cite{Hooper:2010uy}
%\bibitem{Hooper:2010uy}
  D.~Hooper {\it et al.}
  %, J.~I.~Collar, J.~Hall, D.~McKinsey and C.~Kelso,
  %``A Consistent Dark Matter Interpretation For CoGeNT and DAMA/LIBRA,''
  arXiv:1007.1005 [hep-ph].
  %%CITATION = ARXIV:1007.1005;%%

\bibitem{CRESST}
see talk of W.~Seidel at IDM2010.

%\bibitem{lowmassDMbound}
%\cite{Aprile:2010um}
\bibitem{Aprile:2010um}
  E.~Aprile {\it et al.}  [XENON100 Collaboration],
  %``First Dark Matter Results from the XENON100 Experiment,''
  arXiv:1005.0380 [astro-ph.CO].
  %%CITATION = ARXIV:1005.0380;%%

  %\cite{Akerib:2010pv}
\bibitem{Akerib:2010pv}
  D.~S.~Akerib {\it et al.}  [CDMS Collab.~],
  %``A low-threshold analysis of CDMS shallow-site data,''
  arXiv:1010.4290 [astro-ph.CO];
  %%CITATION = ARXIV:1010.4290;%%
  %\cite{Ahmed:2010wy}
%\bibitem{Ahmed:2010wy}
  Z.~Ahmed {\it et al.} [ CDMS-II Collaboration ],
  %``Results from a Low-Energy Analysis of the CDMS II Germanium Data,''
  %Submitted to: Phys.Rev.Lett..
  [arXiv:1011.2482 [astro-ph.CO]].


  %\cite{Savage:2010tg}
\bibitem{Savage:2010tg}
  C.~Savage {\it et al.}
  %, G.~Gelmini, P.~Gondolo and K.~Freese,
  %``XENON10/100 dark matter constraints in comparison with CoGeNT and DAMA:
  %examining the Leff dependence,''
  arXiv:1006.0972 [astro-ph.CO].
  %%CITATION = ARXIV:1006.0972;%%

  %\cite{Sorensen:2010hv}
\bibitem{Sorensen:2010hv}
  P.~Sorensen {\it et al.}
  %, J.~Angle, E.~Aprile, F.~Arneodo, L.~Baudis, A.~Bernstein, A.~Bolozdynya, P.~Brusov {\it et al.},
  %``Lowering the low-energy threshold of xenon detectors,''
  [arXiv:1011.6439 [astro-ph.IM]].




\bibitem{lowmassquestions}
%\cite{Collar:2010gg}
%\bibitem{Collar:2010gg}
  J.~I.~Collar and D.~N.~McKinsey,
  %``Comments on 'First Dark Matter Results from the XENON100 Experiment',''
  arXiv:1005.0838 [astro-ph.CO];
  %%CITATION = ARXIV:1005.0838;%%
%\cite{Collaboration:2010er}
%\bibitem{Collaboration:2010er}
  The XENON Collab.~,
  %``Reply to the Comments on the XENON100 First Dark Matter Results,''
  arXiv:1005.2615 [astro-ph.CO];
  %%CITATION = ARXIV:1005.2615;%%
  %\cite{Collar:2010gd}
%\bibitem{Collar:2010gd}
  J.~I.~Collar and D.~N.~McKinsey,
  %``Response to arXiv:1005.2615,''
  arXiv:1005.3723 [astro-ph.CO];
  %%CITATION = ARXIV:1005.3723;%%
%\cite{Collar:2010nx}
%\bibitem{Collar:2010nx}
  J.~I.~Collar,
  %``Comments on arXiv:1006.0972 'XENON10/100 dark matter constraints in
  %comparison with CoGeNT and DAMA: examining the Leff dependence',''
  arXiv:1006.2031 [astro-ph.CO],
  %%CITATION = ARXIV:1006.2031;%%
  %\cite{Collar:2010ht}
%\bibitem{Collar:2010ht}
%  J.~I.~Collar,
  %``Light WIMP Searches: The Effect of the Uncertainty in Recoil Energy Scale
  %and Quenching Factor,''
  arXiv:1010.5187 [astro-ph.IM];
  %%CITATION = ARXIV:1010.5187;%%
  %\cite{Aprile:2011hx}
%\bibitem{Aprile:2011hx}
  E.~Aprile {\it et al.}  [XENON100 Collab.~],
  %``Likelihood Approach to the First Dark Matter Results from XENON100,''
  arXiv:1103.0303 [hep-ex].
  %%CITATION = ARXIV:1103.0303;%%

%\cite{Feng:2008qn}
\bibitem{Feng:2008qn}
%\cite{Hooper:2008cf}
%\bibitem{Hooper:2008cf}
  D.~Hooper {\it et al.},
  %``The New DAMA Dark-Matter Window and Energetic-Neutrino Searches,''
  Phys.\ Rev.\  {\bf D79}, 015010 (2009).
  [arXiv:0808.2464 [hep-ph]];
  J.~L.~Feng {\it et al.}
  %, J.~Kumar, J.~Learned, L.~E.~Strigari,
  %``Testing the Dark Matter Interpretation of the DAMA/LIBRA Result with Super-Kamiokande,''
  JCAP {\bf 0901}, 032 (2009).
  [arXiv:0808.4151 [hep-ph]];
  %\cite{Niro:2009mw}
%\bibitem{Niro:2009mw}
  V.~Niro {\it et al.}
  %, A.~Bottino, N.~Fornengo and S.~Scopel,
  %``Investigating light neutralinos at neutrino telescopes,''
  Phys.\ Rev.\  D {\bf 80}, 095019 (2009)
  [arXiv:0909.2348 [hep-ph]].
  %%CITATION = PHRVA,D80,095019;%%



%\cite{Savage:2008er}
\bibitem{Savage:2008er}
  C.~Savage {\it et al.}
  %, G.~Gelmini, P.~Gondolo, K.~Freese,
  %``Compatibility of DAMA/LIBRA dark matter detection with other searches,''
  JCAP {\bf 0904}, 010 (2009).
  [arXiv:0808.3607 [astro-ph]].

%\cite{Chang:2010yk}
\bibitem{Chang:2010yk}
  S.~Chang {\it et al.}
  %, J.~Liu, A.~Pierce, N.~Weiner and I.~Yavin,
  %``CoGeNT Interpretations,''
  JCAP {\bf 1008}, 018 (2010)
  [arXiv:1004.0697 [hep-ph]].
  %%CITATION = JCAPA,1008,018;%%

%\cite{Feng:2011vu}
\bibitem{Feng:2011vu}
  J.~L.~Feng {\it et al.}
  %, J.~Kumar, D.~Marfatia, D.~Sanford,
  %``Isospin-Violating Dark Matter,''
  [arXiv:1102.4331 [hep-ph]].

%\cite{Zentner:2009is}
\bibitem{Zentner:2009is}
  A.~R.~Zentner,
  %``High-Energy Neutrinos From Dark Matter Particle Self-Capture Within the
  %Sun,''
  Phys.\ Rev.\  D {\bf 80}, 063501 (2009)
  [arXiv:0907.3448 [astro-ph.HE]].
  %%CITATION = PHRVA,D80,063501;%%

\bibitem{LBNE}
LBNE Conceptual Design Report,
{\tt http://www.phy.bnl.gov/~bviren/lbne/cdr/v4ch13.pdf}

%\cite{Hooper:2010mq}
\bibitem{Hooper:2010mq}
  D.~Hooper, L.~Goodenough,
  %``Dark Matter Annihilation in The Galactic Center As Seen by the Fermi Gamma Ray Space Telescope,''
  Phys.\ Lett.\  {\bf B697}, 412-428 (2011)
  [arXiv:1010.2752 [hep-ph]].




\end{thebibliography}
\end{document}